# Bear Markets and Recessions versus Bull Markets and Expansions

Abdulnasser Hatemi-J

Department of Accounting and Finance, College of Business and Economics, UAE University

Email: AHatemi@uaeu.ac.ae

http://arxiv.org/abs/2009.01343

Abstract

This paper examines the dynamic interaction between falling and rising markets for both the real and the financial sectors of the world's largest economy using asymmetric causality tests. These tests require that each underlying variable in the model be transformed into partial sums of the positive and negative components. The positive components represent the rising markets and the negative components embody the falling markets. The sample period covers some part of the COVID-19 pandemic. Since the data is non-normal and the volatility is time varying, the bootstrap simulations with leverage adjustments are used in order to create reliable critical values when causality tests are conducted. The results of the asymmetric causality tests disclose that the bear markets are causing the recessions as well as the bull markets are causing the economic expansions. The causal effect of bull markets on economic expansions is higher compared to the causal effect of bear markets on economic recessions. In addition, it is found that economic expansions cause bull markets but recessions do not cause bear markets. Thus, the policies that remedy the falling financial markets can also help the economy when it is in a recession.

1. **Introduction**

According to the National Bureau of Economic Research (NBER) a recession is defined as two or more successive quarters of decline in the real GDP in the US. Likewise, an expansion takes place when the real GDP grows for two or more succeeding quarters, moving from a trough to a peak. Based on historical US data, recessions last within an interval of six to eighteen months. A bear market is prevalent when the financial market experiences extended price declines. Usually, it defines a condition in which asset prices fall twenty percent or more from its recent highs due to negative investor sentiment and/or expectations. On the other hand, a bull market represents a situation in which the asset prices are increasing or are expected to increase.[1] The real and financial sectors are anticipated to be closely interlinked. Economic conditions can affect the market conditions and *vice versa*. Stock markets trends can precede economic ones and the other way around. In the literature, the main emphasis is on the relationship between economic growth and the stock market without directly accounting for the potential asymmetric impacts between the two underlying variables. This is the gap in the literature that the current paper intends to fill by making use of the asymmetric causality tests, which explicitly separate the casual impact of the positive changes from the negative ones. Both recessions and bear markets take place usually during the same period more or less. Likewise, expansions and bull markets follow each other. The main question within this context is; what is causing what? Does a recession cause a bear market? Is the causal impact the other way around or do these two variables reinforce each other? Similar questions can be asked pertinent to the potential relationships between expansions and bull markets. In addition, it is important for both investors and policy makers to figure out whether the potential causal impact between the two variables is stronger when markets are rising or falling. The current paper seeks finding empirical answers to these important questions for the world's largest economy and the world's largest financial market.

The reminder of this paper is organized as the following. Section 2 provides a brief literature review. Section 3 describes the data and the methodology. Section 4 presents the empirical results and the final section offers conclusions.

---

[1] Both bull markets and economic expansions take longer time usually compared to recessions and bear markets.

## 2. Literature Review

The importance of financial sector for the development of economic or real sector goes back to the original contributions of Schumpeter (1911). Goldsmith (1969), Mckinnon (1973), Shaw (1973) and Cole and Shaw (1974) advanced the idea further. Inspired by these influential publications, a lot of research has been devoted to the empirical investigation of the dynamic interaction between financial sector and real sector of the economy. Studies conducted by, *inter alia*, Atje and Jovanovich (1993), Korajczyk (1996) in addition to Levine and Zervos (1998) provide empirical support for a strong and positive relationship between the stock market and the economic growth. Prominent studies conducted by Fama (1981) as well as Fisher and Merton (1984) report a significant and positive relationship between the US stock market and the real economic activity of the economy. According to Fama (1990) the growth rates of production describe more than 40% of the stock return variance in the New York Stock Exchange (NYSE), which is the world's biggest equities-based exchange as measured by the total market value of the listed securities trading. Binswanger (2000) finds empirical support that the relationship between the stock market and production as well as the GDP broke down in the bull market of the US in early 1980s. Pena and Rodriguez (2006) provide empirical evidence that supports a significant relationship between the asset prices and the real economic activity in the US and Canada. Mao and Wu (2010) investigate the relationship between the stock market and the GDP in Australia. Their findings indicate that the relationship between the stock market and the GDP is positive and strong in Australia even during the periods of high economic growth and globalization. Suzuki (2012) shows that the stock market booms in countries experiencing a war are in accordance with a model of asset pricing at equilibrium. Gozgor (2015) explores the causal relationship between the financial sector and the real sector for a sample of 58 countries. According to this study there is a significant causal impact running from the financial sector on the real sector in seven developed countries. Furthermore, it is found that the real sector causes the financial sector in five developed countries and in ten developing countries. Tiwari et al. (2018) conduct an empirical investigation for a long sample period covering years 1801-2016 in the US. Their results indicate that the causal effect of the GDP on the stock prices is stronger than the other way around. Pan and Mishra (2018) offer a review of currently published research on this important topic. A study conducted by Darrat (1999) investigates the impact of financial deepening on the growth in Saudi Arabia, the UAE and Turkey using two measures for financial deepening. The first measure is the ratio of currency to M1 and the second one is the inverse form of the broad money velocity. The empirical finding from this study show that there is a long run relationship between the financial deepening on the economic growth in these countries. Ben Naceur et al. (2008) conducts an empirical investigation for elven countries in the Middle East and North Africa (MENA) on the potential impact of stock market liberalization on the economic performance. They do not find empirical evidence for the stock market liberalization having any significant impact on growth or investment in this panel. Nevertheless, an opposite causal impact is found, which is positive in the long run and negative in the short run. Ostermalm (1996) finds that there is a steady state relationship between the stocks and the GDP in Sweden during the last three decades and the GDP as a fundamental value factor for the stocks has a significant forecasting power on the stock market. However, the mentioned studies do not allow for the asymmetric causal impacts in the empirical investigation. The current paper aims at filling this gap in the literature by conducting asymmetric causality tests for the world's largest economy. Accounting for the asymmetric impacts in the empirical investigation is consistent with the way reality operates because the reaction of the economic agents tend to be situation dependent. That is people tend to react more to negative changes of a variable compared to the positive changes of the same variable.

3. **Methodology**

It is well established in the existing literature that economic actors tend to respond stronger to a negative condition compared to a corresponding positive one. This seems particularly to be the case when it comes to the financial issues. This asymmetric impact cannot not be captured by the standard symmetric methods. Thus, we are making use of asymmetric methods in order to account for this potential unevenness. The asymmetric methods are also more informative because situation specific information can be extracted pertinent to both falling and rising markets. Furthermore, the asymmetric methods can to be more efficient compared to the symmetric ones with regard to model specification because asymmetric methods are non-linear and thereby more general approach. Since financial data tends to be non-normal with time-varying volatility, the asymptotic critical values are not be precise when hypotheses are tested.

In the current paper, data specific critical values are generated via bootstrap simulations in order to enhance the reliability of the empirical findings. To account for the potential asymmetric causal impacts, we implement the tests suggested by Hatemi-J (2012). It is commonly observed in the literature that macroeconomic and financial time series are usually characterized by a unit root. Therefore, the real GDP (denoted by *Y*) can be defined as the following unit root process:

$$Y_t = Y_{t-1} + e_t \qquad (1)$$

for time span *t = 1, 2,..., T*. Where $e_t$ is a white noise an error term. By making use of the recursive approach, the following solution can be found for the data generating process expressed by equation (1):

$$Y_t = Y_0 + \sum_{i=1}^{t} e_i \qquad (2)$$

The constant value of $Y_0$ signifies the initial value of the real GDP. In order to identify the positive and negative components the definitions $e_i^+ = Max(e_i, 0)$ and $e_i^- = Min(e_i, 0)$ are used.[2] This in turn results in having $e_i = e_i^+ + e_i^-$. Therefore, the expression below can stated

$$Y_t = Y_0 + \sum_{i=1}^{t} e_i = Y_0 + \sum_{i=1}^{t} e_i^+ + \sum_{i=1}^{t} e_i^- \qquad (3)$$

The partial sums for positive and negative components of the underlying variable can be expressed $Y_t^+ = \frac{Y_0}{2} + \sum_{i=1}^{t} e_i^+$ and $Y_t^- = \frac{Y_0}{2} + \sum_{i=1}^{t} e_i^-$. Note that the required condition $Y_t = Y_t^+ + Y_t^-$ is fulfilled. The other variable (i.e. the stock market index) can be transformed similarly. These partial components make it operational to conduct the asymmetric causality tests. For dealing with the economic expansions and the bull markets the vector is $Z_t^+ = (Y_t^+, S_t^+)$, which can be used for estimating the following vector autoregressive model of order *l*, VAR(*l*) is estimated.[3]

$$Z_t^+ = B_0^+ + B_1^+ Z_{t-1}^+ + \cdots B_l^+ Z_{t-l}^+ + \varepsilon_t^+ \qquad (4)$$

---

[2] Granger and Yoon (2002) utilized these definitions for implementing tests of hidden cointegration. Hatemi-J (2020) extends this method for testing for hidden panel cointegration.

[3] Note that the vector $Z_t^- = (Y_t^-, S_t^-)$ can be used for capturing the dynamic interaction between the negative components of the underlying variables.

where $B_0^+$ is a *2×1* vector of intercepts and $\varepsilon_t^+$ denotes a *2×1* vector of error terms. The $B_l^+$ is a *2×2* matrix of coefficients to be estimated based on the optimal lag order *l* (*l* = 1, …, *l-max*). The optimal lag order is carefully chosen via minimizing the information criterion expressed below

$$HJC = \ln(|\widehat{\Delta}_l|) + \frac{l[v^2 \ln T + 2v^2 \ln(\ln T)]}{2T}, \qquad l = 1, \cdots, l_{max} \qquad (5)$$

Here $|\widehat{\Delta}_l|$ denotes the determinant of the variance and covariance matrix of $\varepsilon_t^+$, which is estimated based on lag *l* for *v* variables in the VAR model and for the sample size $T$.[4] Note that *l*-max is the maximum lag order that is considered in the estimations. Subsequently, the following hypothesis of non-causality is tested by the Wald test:[5]

$H_0$: The row *j* and column *l* element in $B_l$ is equal to zero for *l* = 1, …, *l-max*. (6)

Before presenting the Wald test, we introduce the following denotations for practical reasons:[6]

$$W^+ = (Z_1^+, \cdots, Z_T^+), \quad (v \times T) \; matrix,$$

$$D^+ = (B_0^+, B_1^+, \cdots, B_k^+), \quad (v \times (1 + v \times l)) \; matrix,$$

$$X_t^+ = \begin{bmatrix} 1 \\ Z_t^+ \\ Z_{t-1}^+ \\ \vdots \\ Z_{t-l-1}^+ \end{bmatrix}, \quad ((1 + v \times l) \times 1) \; matrix,$$

$$W^+ = (W_0^+, \cdots, W_{T-1}^+), \quad ((1 + v \times l) \times T) \; matrix,$$

Furthermore,

$$\varepsilon^+ = (\varepsilon_1^+, \cdots, \varepsilon_T^+), \quad (v \times T) \; matrix,$$

Via these denotations the VAR(*l*) model can be formulated compactly as the following:[7]

---

[4] The information criterion expressed in equation (5) is proposed by Hatemi-J (2003, 2008). The conducted Monte Carlo simulations demonstrates that this information criterion can effectively select the optimal lag length even in situations in which ARCH effects and unit roots prevail. In addition, this information criterion has good forecasting properties based on the simulation results. See also Mustafa and Hatemi-J (2020) for additional details on the properties of this information criterion.

[5] Since the variables is integrated of the first degree, an unrestricted extra lag was added to the VAR model in order to account for the effect of the unit root based on the results of Toda and Yamamoto (1995).

[6] Naturally, this compact formulation is possible if and only if the initial values are available according to Lutkepohl (2005).

[7] The original of the VAR model is the seminal article of Sims (1980).

$$W^+ = D^+X^+ + \varepsilon^+ \tag{7}$$

The null hypothesis can stated as $H_0: C\alpha = 0$. To test the hypothesis the succeeding test statistic based on Wald (1939)'s idea is estimated:

$$Wald = (C\alpha)'\left[C\left((X^{+\prime}X^+)^{-1}\otimes\hat{\Delta}_U\right)C'\right]^{-1}(C\alpha) \tag{8}$$

Notice that $\alpha = vec(D^+)$, where *vec* is the vectorization operator. The denotation $\otimes$ represents the Kronecker product used for creating the needed block matrix of the underlying matrixes. The indicator matric *C* has the dimension *l×v*(1+*v×l*) and it contains elements of ones that correspond to the restricted parameters under the null hypothesis and it has zeros for the unrestricted parameters. $\hat{\Delta}_U$ is the variance and covariance matrix that is estimated for the unrestricted model based on the lag order *l*. The unrestricted model is the one that does not impose the restrictions implied by the null hypothesis. Asymptotically, the Wald test as defined in equation (8) has a $\chi^2$ distribution with *l* degrees of freedom. Nevertheless, if the data is not fulfilling the normality assumption and the volatility is time varying, the critical values based on the asymptotic distribution tend to lack precision. Based on the investigation of historical data, non-normality and time varying volatility appears to be a common characteristic of most financial data. Thus, implementing causality tests based on asymptotical critical values might not result in accurate inference. According to Hacker and Hatemi-J (2006, 2012) making use of bootstrap simulations with leverage adjustments improves on the performance of the causality tests compared to the asymptotic ones. The statistical software components produced in Gauss by Hacker and Hatemi-J (2010) and Hatemi-J (2011) are used for implementing both the symmetric as well as the asymmetric causality tests via bootstrap simulations with leveraged adjustments.

### 4. Data and the Empirical Results

Our sample is on quadratic basis and it covers the period 1960:01:Q1-2020:01:Q1. The real GDP and the total share prices for all shares of the US are collected from the Federal Reserve Economic Data (FRED, 2020) database, which is provided online by the Federal Reserve Bank of St. Louis.

The first step in our empirical investigation is testing for unit roots by making use of the test statistics developed by Ng and Perron (2001), which also determines the optimal lag order in the unit root equation. These test results are presented in Table 1, which indicate that each variable contains one unit root. Thus, it is important to include an unrestricted lag order in the VAR model when tests for causality are conducted in order to account for the effect of the unit root based on the suggestions of Toda and Yamamoto (1995). The next step is to conduct multivariate tests for normality and ARCH effects in the VAR model. The results of these diagnostic tests are presented in Table 2. It is evident from the tests that the residuals in all three VAR models are not normally distributed and the volatility is time varying in each case. This implies that conducting tests for causality based on the asymptotical critical values is not going to be precise. In order to remedy this shortcoming, we make use of the bootstrap simulations with leverage adjustments for producing reliable critical values when tests for causality are implemented.

Table 1. The Unit Root Test Results.

| VARIABLE | Test Value For $H_0$: I(1) | Test Value For $H_0$: I(2) |
|---|---|---|
| $Y$ | 0.92013 | -5.72831 |
| $Y^+$ | 0.93919 | -14.4391 |
| $Y^-$ | 2.15901 | -102.746 |
| $S$ | 1.26655 | -102.521 |
| $S^+$ | 1.59322 | -90.9764 |
| $S^-$ | 1.54933 | -108.458 |

Notes:
1. $S$ is the total share prices for all shares of the US (Index 2015=100). $Y$ is representing the real GDP. The sample period starts with 1960-01-Q1 and ends with 2020-01-Q1. The vector ($S^+$, $Y^+$) denotes the partial cumulative sums for the positive components and ($S^-$, $Y^-$) signifies the partial cumulative sums for the negative components.
2. Ng and Perron (2001) test is used. The critical values are -13.80, -8.10 and -5.70 at the 1%, 5% and 10% significance levels respectively.

Table 2. The Results of Multivariate Normality and Multivariate ARCH Tests.

| VARIABLE IN THE VAR MODEL | P-Value For $H_0$: Multivariate Normality | P-Value For $H_0$: No Multivariate ARCH |
|---|---|---|
| $(Y, S)$ | <0.00001 | 0.00020 |
| $(Y^+, S^+)$ | <0.00001 | 0.00010 |
| $(Y^-, S^-)$ | <0.00001 | <0.00001 |

Notes: The test method developed by Doornik and Hansen (2008) is used for testing the null hypothesis of multivariate normality. The test statistic suggested by Hacker and Hatemi-J (2005) is applied for testing the null hypothesis of no multivariate ARCH effects. This test is available as automatic procedure in econometric software entitled Regression Analysis of Time Series (RATS) produced by Estima (2020).

The test results of the symmetric and asymmetric causality tests based on bootstrap simulations and with leverage corrections are presented in Table 3. Based on the symmetric causality tests results we can conclude that the causality is running from the stock market to the real GDP and there is no reverse causality. The causal impact of the stock market on the real economic activity is positive. The asymmetric causality tests reveal that the bear markets are causing recessions and the bull markets are causing economic expansions. The causal impact of bull markets on economic expansions is stronger than the causal impact of bear markets on recessions according to the estimated causal parameters.. Furthermore, the causality test results reveal that economic expansions cause bull markets positively but there is no significant causality running from recessions on the bear markets. The implications of this might be that the policies designed for remedying the falling markets are going to support the performance of the economy when it is in a recession state.

Table 3. The Symmetric and Asymmetric Causality Test Results

| NULL HYOTHESIS | Test Value | Bootstrap CV at 1% | Bootstrap CV at 5% | Bootstrap CV at 10% | Causal Parameter | Lag Order |
|---|---|---|---|---|---|---|
| $S \not\Rightarrow Y$ | 12.081*** | 11.836 | 7.872 | 6.358 | 0.0055 | 3 |
| $S^+ \not\Rightarrow Y^+$ | 8.707** | 12.167 | 8.157 | 6.379 | 0.0130 | 3 |
| $S^- \not\Rightarrow Y^-$ | 15.345** | 24.468 | 8.007 | 4.583 | 0.0003 | 2 |
| $Y \not\Rightarrow S$ | 6.213 | 12.069 | 8.129 | 6.441 | | 3 |
| $Y^+ \not\Rightarrow S^+$ | 11.280** | 11.782 | 8.167 | 6.411 | 0.0341 | 3 |
| $Y^- \not\Rightarrow S^-$ | 1.689 | 23.341 | 7.568 | 4.355 | | 2 |

Notes:

1. $S$ is the total share prices for all shares of the US and $Y$ is representing the real GDP. The vector $(S^+, Y^+)$ denotes the partial cumulative sum for positive components and $(S^-, Y^-)$ signifies the partial cumulative sum for negative components.
2. The denotation *** means significant at 1% level and ** means significant at 5% level.
3. One additional unrestricted lag was included in each VAR model in order to account for the unit root effect based on the recommendations by Toda and Yamamoto (1995).

## 5. Conclusions

When the economy plunges into recession the stock markets tend to fall and vise versa. Similar relationship is expected to be observed between the booming economy and the rising financial markets. Measuring the potential relationship between these two crucial variables and finding out what drives what is of fundamental importance to both investors and policy makers. The current paper investigates empirically this issue for the world's largest economy via asymmetric causality tests. Quadratic data is used for the period 1960:01:Q1-2020:01:Q1, which covers some part of the fastest and the steepest recession in history due to the unexpected COVID-19 pandemic. However, the economy is starting to recover due to impressive federal aid in the form of stimulus checks, substantial unemployment benefits and forgivable loans to the small business organizations according to Petras and Davidson (2020).

We make use of the asymmetric causality tests because these methods reflect better the way in which markets operate in reality. It is widely established in the literature that investors react stronger to the negative conditions compared to the positive ones. This important aspect cannot be covered by the symmetric methods. In addition, asymmetric methods are more informative and useful in the sense of providing situation specific information pertinent to both falling and rising markets. Furthermore, the asymmetric methods are likely to be more efficient in comparison to the symmetric ones from the point of model specification. This is the case because asymmetric methods can capture some form of non-linearity in the relationship between the variables, which can be considered as a more general framework for conducting the empirical investigation. Another issue that needs to pay attention to is whether the desirable statistical assumptions for a good model are fulfilled. Based on the conducted diagnostic tests our data appears to be non-normal with ARCH effects. In such cases, the causality tests via the asymptotical critical values might not provide precise reference. One solution to this problem is to make use of the bootstrap simulations with leverage adjustments in order to create reliable critical values.

The results from both symmetric and asymmetric causality tests via bootstrap simulations and with leverage improvements reveal the following. The symmetric causality tests show that there is a positive casual impact running from the stock market to the real GDP and not vice versa. The asymmetric causality tests on the other hand indicate that the bear markets are causal factors for recessions and the bull markets are causal factors for economic expansions. The estimated casual parameters disclose that the causal effect of bull markets on economic expansions is higher than the causal impact of bear markets on recessions. Finally, the test results indicate that economic expansions are causing the bull markets; however, recessions do not cause bear markets. These empirical findings imply that any policy that supports the financial market when it is falling can also support the economy when it is in a recession.

**Data Availability Statement**

The data used in the empirical investigation is available online from the following source:

FRED (2020) The Federal Reserve Bank of St. Louis. Link: https://fred.stlouisfed.org/